\documentclass[twocolumn,footinbib,floatfix,aps,pra]{revtex4-1}
\usepackage[usenames,dvipsnames]{xcolor} 
\usepackage[normalem]{ulem}

\usepackage{amsmath,amssymb}
\usepackage{amsfonts}
\usepackage{bm}
\usepackage{bbm}
\usepackage{color}
\usepackage{latexsym}
\usepackage[english]{babel}
\usepackage{times}
\usepackage{stmaryrd}
\usepackage{psfrag,graphicx}
\usepackage{subfigure}
\usepackage{appendix}
\usepackage{enumitem}
\usepackage{bbold}

\def\CD{\textcolor{black}}

\definecolor{mygrey}{gray}{0.35}
\definecolor{myblue}{rgb}{0.2,0.2,0.8}
\definecolor{myzard}{cmyk}{0,0,0.05,0}
\definecolor{mywhite}{rgb}{1,1,1}
\definecolor{myred}{rgb}{0.9,0.1,0.}
\usepackage[colorlinks=true,citecolor=myblue,linkcolor=myred,urlcolor=myblue]{hyperref}

\newcommand{\dg}{\dagger}
\newcommand{\ket}[1]{\left| #1 \right>} 
\let\baraccent=\= 

\newcommand{\half}[1][1]{\frac{#1}{2}} 

\newcommand{\be}{\begin{equation}}
\newcommand{\ee}{\end{equation}}
\newcommand{\ben}{\begin{eqnarray}}
\newcommand{\een}{\end{eqnarray}}
\newcommand{\bes}{\begin{subequations}}
\newcommand{\ees}{\end{subequations}}
\newcommand{\bF}{\begin{figure}}
\newcommand{\eF}{\end{figure}}

\newcommand{\trace}[1]{\mathrm{Tr}\left(#1\right)}

\newcommand{\DBcolor}{teal}
\newcommand{\DB}[2][]{\ifx\\#1\\\textcolor{\DBcolor}{#2}\else\textcolor{\DBcolor}{\textsc{#2}}\fi}

\begin{document}


\title{Gaussian systems for quantum-enhanced multiple phase estimation}

\author{Christos N. Gagatsos}
\affiliation{Department of Physics, University of Warwick, Coventry CV4 7AL, United Kingdom}

\author{Dominic Branford}
\affiliation{Department of Physics, University of Warwick, Coventry CV4 7AL, United Kingdom}

\author{Animesh Datta}
\affiliation{Department of Physics, University of Warwick, Coventry CV4 7AL, United Kingdom}

\date{\today}


\begin{abstract}

For a fixed average energy, the simultaneous estimation of multiple phases can provide a better total precision than estimating them individually. We show this for a multimode interferometer with a phase in each mode, using Gaussian inputs and passive elements, by calculating the covariance matrix. The quantum Cram\'{e}r-Rao bound provides a lower bound to the covariance matrix via the quantum Fisher information matrix, whose elements we derive to be the covariances of the photon numbers across the modes. We prove that this bound can be saturated. In spite of the Gaussian nature of the problem, the calculation of non-Gaussian integrals is required, which we accomplish analytically. We find our simultaneous strategy to yield no more than a factor-of-$2$ improvement in total precision, possibly because of a fundamental performance limitation of Gaussian states. Our work shows that no modal entanglement is necessary for simultaneous quantum-enhanced estimation of multiple phases.

\end{abstract}

\maketitle

\section{Introduction}  
Parameter estimation with quantum-enhanced precision has the potential to provide substantial technological advances as well as deep insights into the fundamental workings of Nature.
Originating in the quest for the increased sensitivity requirements for detecting gravitational waves using laser interferometers with squeezed light~\cite{Caves:1981aa,Adhikari2014}, the field now encompasses a variety of scenarios studying the quantum limits of sensing~\cite{PARIS:2009aa,Giovannetti:2011aa,Demkowicz-Dobrzanski:2015aa,Toth:2014aa}.
Relative phase estimation in a two-mode interferometer is by far the most common, although some attention has also been cast to the simultaneous estimation of multiple parameters at the quantum limit~\cite{YUEN:1973aa,FUJIWARA:1994aa,Genoni:2013aa,Humphreys:2013aa,Crowley:2014aa,Yao2014b,Knott2016}.

A fundamental bound on the precision of an estimation is the quantum limit on the variance of the estimator. This is set by the quantum Cram\'er-Rao bound (QCRB)~\cite{BRAUNSTEIN:1994aa} and valuable insights into the working of quantum mechanics have been obtained by studying it in the multiparameter scenario~\cite{Gill:2000aa, Vidrighin:2014aa, Berry:2015aa}.
In addition to this fundamental understanding, several scenarios of practical and technological interest are intrinsically multiparameter estimation problems, leading to new methodologies of obtaining quantum enhancements arising purely from the multidimensional nature of the problem.
This includes magnetic-field sensing in three dimensions~\cite{Baumgratz:2016aa} and imaging~\cite{Humphreys:2013aa,Yao2014,Yue2014,Liberman2015}.
These proposed schemes use a fixed number of photons in multimode entangled states, which are not easy to prepare for increasing photon numbers.

Concerning Gaussian states and their role in estimation theory, general expressions have been derived which are useful for evaluating the quantum Fisher information matrix~\cite{Monras:2010aa,Monras:2011aa,Monras2013, Gao:2014aa, Safranek2015, Pirandola2015} but the explicit expressions found in these works are limited to two-parameter estimation problems.
Reference~\cite{ZhangFan2014} utilises the quantum Ziv-Zakai bound~\cite{Tsang2012} to numerically study the precision limit of up to $16$-mode squeezed vacuum states and find in their example an improvement with simultaneous strategies without  quantifying the factor of the improvement.

\begin{figure}
\includegraphics[width=\linewidth]{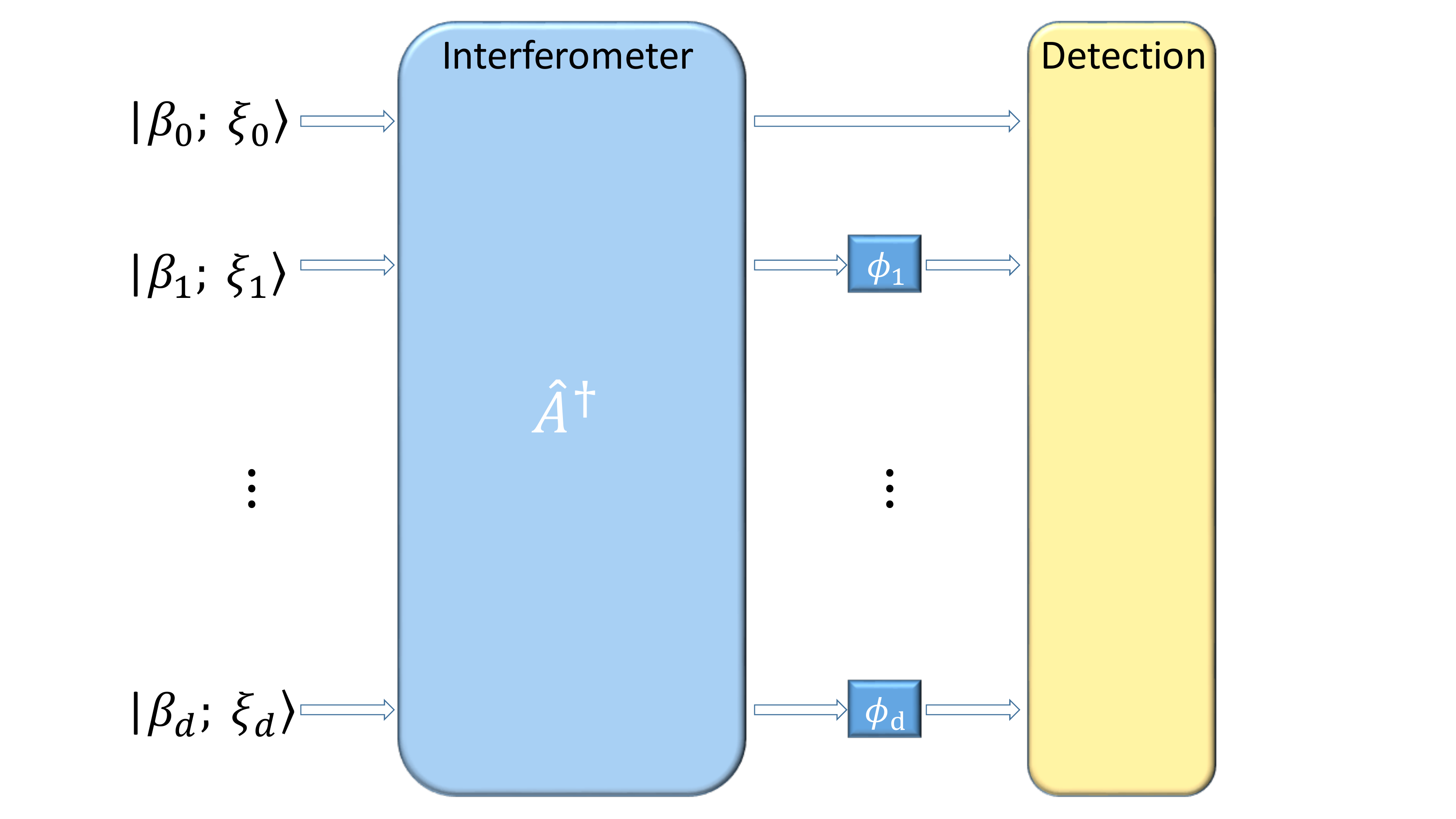}
\caption{A $d+1$ mode interferometer with the most general pure Gaussian input, produced from a general pure and separable Gaussian state followed by a passive element. The resultant state undergoes the phase shifts to be estimated.}
\label{schematic1}

\includegraphics[width=\linewidth]{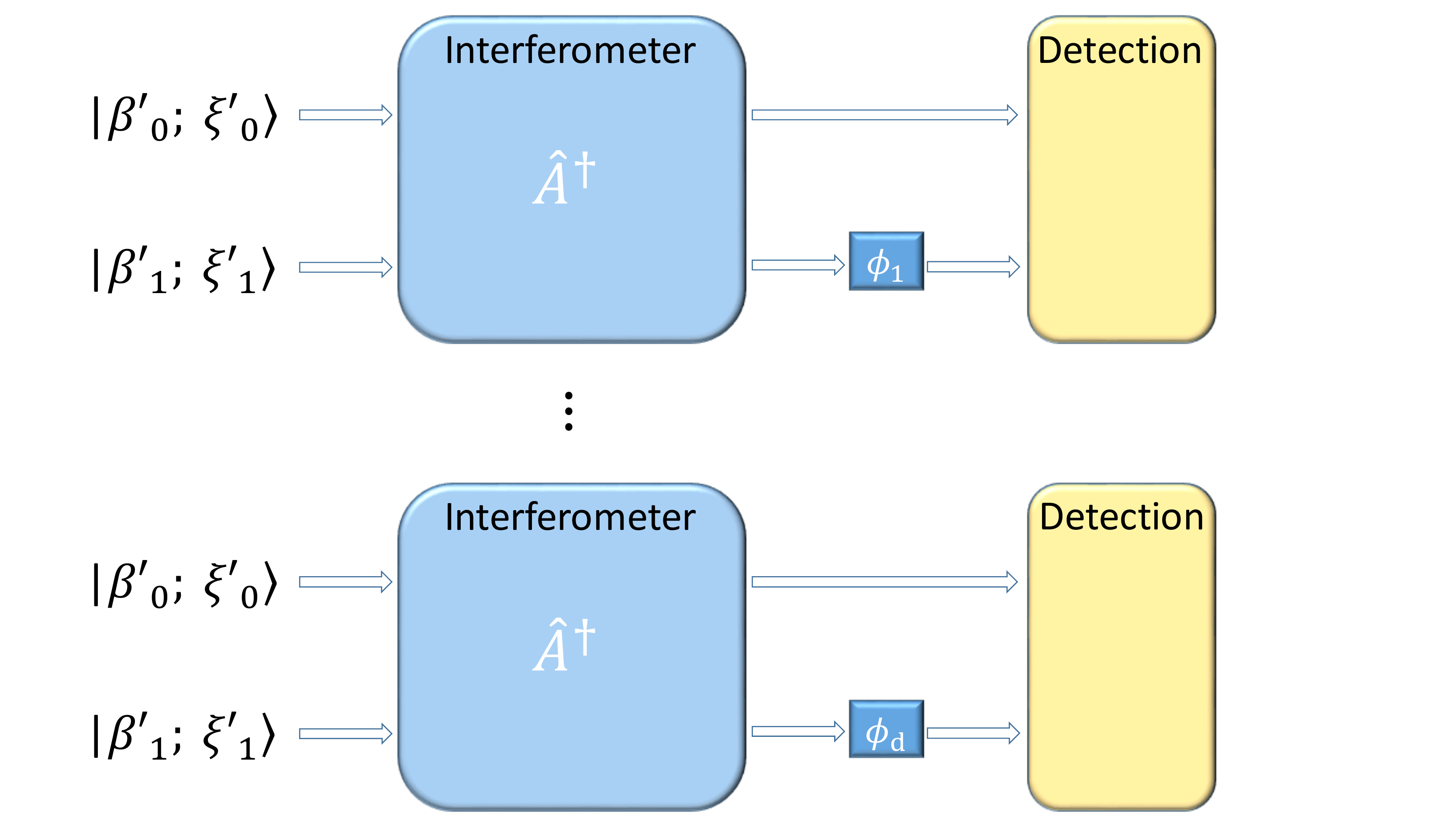}
\caption{A two-mode interferometer with the most general pure Gaussian input, produced from a general pure and separable Gaussian state followed by a passive element. The resultant state undergoes the phase shift to be estimated. This is repeated $d$ times, i.e. for the number of phases to be estimated or it can be viewed as $d$ parallel, individual estimations. The total energy for simultaneous and individual estimation is the same.}
\label{schematic2}
\end{figure}

In this work we show that for an arbitrary number of phases and a fixed \emph{average} amount of energy, simultaneous estimation (Fig.~\ref{schematic1}) of a fixed number of phase parameters is better than individual estimation (Fig.~\ref{schematic2}). We do so by obtaining an analytical expression for the quantum Fisher information matrix (QFIM) as a function of the number of phases and the total average energy. 
In spite of the improvement found in the simultaneous case, we observe that under assumptions of equal magnitude squeezing in each mode and a multimode interferometer which is an orthogonal transform presents at most a factor-of-2 improvement, pointing to potential limitations of Gaussian states in multiparameter quantum metrology.

The QFIM bounds the covariance matrix for multiple phase estimation, and our results are derived for pure Gaussian states in terms of the Husimi $Q$ function.
Gaussian states are easier to prepare in practice than fixed particle number states, and while couched in the language of optical systems, our work also applies to bosonic degrees of freedom of matter systems.
We show that our bounds are attainable, and discuss the implications of the factor-of-2 improvement.

Our work may thus improve the performance of optical techniques in quantum imaging~\cite{Kolobov2000} and possibly gravitational wave astronomy~\cite{GravitationalWaveInternationalCommittee2010}, as well as optomechanical systems employed in fundamental studies~\cite{Arvanitaki2013,Plato2016}. 
Some of these have been studied experimentally in quantum optics, where noise reduction has been observed using correlated photon pairs~\cite{Brida2010} and multimode squeezed light~\cite{Embrey2015}. More interesting is the constant amount of improvement possible,  unlike the fixed \emph{peak} energy scenario \cite{Humphreys:2013aa} where the improvement scales linearly with the number of parameters. While the limited quantum information processing capabilities of Gaussian states have long been recognised in computation and communication \cite{Barlett2002, Eisert2002, Fiurasek2002, Giedke2002}, ours is a possible instance in quantum metrology.
It is interesting to note that this facet of quantum metrology only appears at the multiple phase level, since Gaussian states are known to achieve the full potential of quantum-enhanced single phase estimation~\cite{Lang2014}.

The paper is organized as follows. In Sec.~\ref{sec:phase} we define the phase shifting, the simultaneous and the individual estimation scenarios. In Sec.~\ref{sec:bounds} we discuss the Cram\'er-Rao bound and its attainability. In Sec.~\ref{sec:qfim} we calculate analytically, under the assumptions we do later on, the QFIM and the trace of its inverse  and in Sec.~\ref{sec:simVSind} we proceed with the comparison of the simultaneous and individual scenarios. Finally, in Sec.~\ref{sec:conclusions} we wrap up and discuss our findings.

\section{phase estimation setup}
\label{sec:phase} 

We study the quantum-limited estimation of $d$ phases $\bm{\phi} \in \mathbb{R}^d$ using a $d+1-$mode pure quantum probe state $\ket{\Psi}$, as shown in Fig.~\ref{schematic1}.
The state $\ket{\Psi}$ picks up the phases $\bm{\phi}$ via $|\Phi \rangle = \hat{\mathcal{U}}_{\bm{\phi}} |\Psi \rangle$~\footnote{In what follows, calligraphic and/or hatted alphabets such as $\mathcal{\hat{A}},\ \mathcal{\hat{U}}, \hat{n}$ denote operators, while bold upper-case alphabets $\mathbf{A},\ \mathbf{U}$ denote their matrix representations (or other matrices, depending on the context).
Letters with two indices such as $A_{i,j},\ U_{i,j},\ h_{i,j}$ represent matrix elements.
Bold lower-case characters denote vectors such as $\bm{\phi}$.}.
The parameters to be estimated are encapsulated in
\begin{align}
\nonumber \hat{\mathcal{U}}_{\bm{\phi}} & = \hat{\mathcal{U}}'_{\bm{\phi}} \exp(-i \varphi \hat{n}_0)\\
\nonumber & =\exp(i \phi_1 (\hat{n}_1-\hat{n}_0) + \ldots + i \phi_d (\hat{n}_d-\hat{n}_0) )\\
& = \exp\left(i \sum_{i=1}^d \phi_i (\hat{n}_i-\hat{n}_0) \right) = \exp(i \bm{\phi}\hat{\bm{g}}),
\label{PhiSU}
\end{align}
where the unitary operator $\hat{\mathcal{U}}'_{\bm{\phi}}=\exp(i \phi_0 \hat{n}_0 +i \phi_1 \hat{n}_1  + \ldots + i \phi_d \hat{n}_d ),$ $\varphi = \phi_0 + \ldots + \phi_d $ captures an unmeasurable overall phase, $\bm{\phi} \equiv (\phi_1, \ldots, \phi_d)^T$ and $\hat{g}_i=\hat{n}_i-\hat{n}_0$ are the generators.
The $\hat{g}_i$ are traceless and Hermitian, as $SU(n)$ generators ought to be.
Indeed, our problem is a special case of $SU(n)$ interferometry, with the parameters to be estimated restricted to a diagonal subgroup.
The reduction of the phase-encoding unitary from an element of the unitary group to an element of the special unitary group is therefore tantamount to accounting for the unmeasurable (global) phase $\phi$.
The nuanced role of a reference mode in quantum interferometry was recently addressed in Ref.~\cite{Jarzyna2012}.

The input $\ket{\Psi}$ is taken to be a pure Gaussian state---the outcome of the interaction of $d+1$ coherent squeezed states with a passive multimode quantum optical element $\mathcal{\hat{A}}^{\dg}$ via $|\Psi \rangle = \mathcal{\hat{A}}^{\dg} \prod_{k=0}^{d} |\beta_k;\xi_k\rangle, $ where the squeezings $\xi_k=|\xi_k| e^{i \theta_k},$ and displacements $\beta_k$ are introduced through the corresponding operators as $\hat{D}(\beta_k)\hat{S}(\xi_k)|0\rangle = |\beta_k;\xi_k\rangle$.
We are able to make the choice of complex displacements and positive squeezings without loss of generality~\footnote{\CD{A local rotation $\mathcal{\hat{U}}_k(\omega_k)$ acting on a squeezed state $| \gamma_k, |\xi_k| e^{i \theta_k} \rangle$ gives $\mathcal{\hat{U}}_k(\omega_k) | \gamma_k, |\xi_k| e^{i \theta_k} \rangle = | \gamma_k e^{-i \omega_k}, |\xi_k| e^{i (\theta_k - \omega_k/2)} \rangle$. By choosing $\omega_k=2 \theta_k$ we get $| \beta_k, |\xi_k|  \rangle$, where $\beta_k = \gamma_k e^{-2 i \theta_k}$ (note that $\beta_k$ and $\gamma_k$ are left to be arbitrary complex numbers). The conjugate transpose of the local rotations $\mathcal{\hat{U}}_k(2 \theta_k)$ can be considered as part of the the general transformation $\mathcal{\hat{A}}^\dagger$ which will now act on the squeezed states  $| \beta_k, |\xi_k|  \rangle$ to still give a general pure Gaussian state.}} to still obtain a general pure Gaussian state~\cite{Braunstein:2005aa}.

Such a state has an average energy of $|\beta_k|^2+\sinh^2|\xi_k|$ in mode $k,$ and our aim is to compare individual and simultaneous estimation 
strategies for $\bm{\phi}$ using the same average input energy totalled over all the modes.
The restriction to squeezed states is primarily motivated by the relative ease of production and manipulation in the laboratory, and demonstration of their relevance in studies of, for instance, quantum information science~\cite{Cerf2007} and gravitational wave astronomy~\cite{Schnabel2010}.
It also avoids, for a fixed mean, the possibility of unbounded variance in particle number. It is not to ease our analytical calculations, as we explain later.

\section{Bounds on precision of estimation}
\label{sec:bounds} 

The performance of any estimation process is captured by the covariance matrix $\mathbf{V}(\bm{\phi})$, the covariance of the estimators for unbiased estimators. This is lower bounded as
\begin{equation}
\label{eq:qcrb}
\mathbf{V}(\bm{\phi}) \geq \mathbf{H}^{-1},
\end{equation}
according to the quantum Cram\'er-Rao bound, where $\mathbf{H}$ is the quantum Fisher information matrix (QFIM)~\cite{Helstrom:1976aa,PARIS:2009aa}.
Equation (\ref{eq:qcrb}) is a matrix inequality, meaning $\mathbf{V}(\bm{\phi}) - \mathbf{H}^{-1}$ is positive semidefinite. The QFIM $\mathbf{H}$ is a real, positive definite, symmetric matrix.
The QFIM can be written in terms of the symmetric logarithmic derivatives (SLDs); $\hat{\mathcal{L}}_i$ for the phases $\phi_i$, are given by 
\begin{equation}
\frac{\partial \hat{\rho}_{\bm{\phi}}}{\partial \phi_i} = \frac{\hat{\mathcal{L}}_i \hat{\rho}_{\bm{\phi}}+\hat{\rho}_{\bm{\phi}} \hat{\mathcal{L}}_i}{2}.
\label{SLDs}
\end{equation}
The QFIM is then $H_{i,j}=\textrm{Tr}\left(\hat{\rho}_{\bm{\phi}}\left(\hat{\mathcal{L}}_i\hat{\mathcal{L}}_j+\hat{\mathcal{L}}_j\hat{\mathcal{L}}_i\right)\right)/2$.

The saturation of the quantum Cram\'er-Rao bound is a two step procedure: equalities in both the classical and quantum Cram\'{e}r-Rao bounds are required.
The former equality requires the use of an efficient, unbiased estimator~\cite{BRAUNSTEIN:1994aa}, in the asymptotic limit maximum likelihood is such an estimator and a convergence to this limit can typically be obtained in a reasonable number of trials~\cite{Braunstein1992,Blandino2012}

The attainability of the latter (quantum) equality is satisfied if the SLDs commute, this is sufficient to prove the existence of a saturating positive operator-valued measure (POVM), the common eigenbasis of the SLDs.
In the case of single parameter estimation the existence of a saturating POVM is therefore trivial.
A looser condition for the attainability of the latter (quantum) limit with pure states is~\cite{Matsumoto:2002aa, Szczykulska2016}
\begin{align} 
\langle \Phi | \left[\hat{\mathcal{L}}_i,\hat{\mathcal{L}}_j\right] | \Phi \rangle = \textrm{Tr} \left( \hat{\rho}_{\bm{\phi}} \left[\hat{\mathcal{L}}_i,\hat{\mathcal{L}}_j\right]  \right) = 0.
\label{condition}
\end{align}
For commuting generators,
\begin{equation}
\nonumber\frac{\partial \hat{\rho}_{\bm{\phi}}}{\partial \phi_j} =i \left[\hat{g}_j,\hat{\rho}_{\bm{\phi}} \right],
\end{equation}
we find (with pure state probes) $\hat{\mathcal{L}}_j = 2 i \left[ \hat{g}_j,\hat{\rho}_{\bm{\phi}} \right]$. 
Using the fact that the generators commute, the cyclicity of the trace and purity of the probe states $\ket{\Psi},$ it is easy to show that the condition in Eq. (\ref{condition}) is satisfied.

\section{Computation of the QFIM}\label{sec:qfim} For any pure state $|\Phi\rangle,$ the QFIM reads $H_{i,j}=4 \Re \left(\langle \partial_i \Phi| \partial_j \Phi \rangle - \langle \partial_i \Phi | \Phi \rangle \langle \Phi | \partial_j \Phi \rangle \right),$ where $\Re(\cdot)$ denotes the real part and $|\partial_i \Phi \rangle \equiv \left( \partial  /\partial \phi_i\right) |\Phi \rangle$. 
For $d$ phase parameters and the corresponding phase shift generators $\{\hat{g}_i\}$, the $d\times d$ QFIM reduces to~\cite{Baumgratz:2016aa}

\begin{align}
\nonumber H_{i,j} &= 4 ( \langle \hat{g}_i \hat{g}_j \rangle - \langle \hat{g}_i \rangle \langle \hat{g}_j \rangle)\\
& =  4 (h_{i,j}-h_{i,0}-h_{0,j}+h_{0,0}),
\label{FisherInfoVar}
\end{align}
with $h_{i,j} = \langle \hat{n}_i \hat{n}_j \rangle - \langle \hat{n}_i \rangle \langle \hat{n}_j \rangle$. The expectation values are calculated for the initial state $\ket{\Psi}.$ Note that for the matrix elements $H_{i,j}$ the indices $i,j$ run from $1$ to $d$, while for the matrix elements $h_{i,j}$ the indices $i,j$ run from $0$ to $d$. Note that $h_{i,0},\ h_{0,j}$ and $h_{0,0}$ give rise to rank-$1$ matrices, and therefore the QFIM can be inverted using the Sherman-Morrison formula. 

We use the Husimi $Q$ representation to calculate the expectation values in Eq.~(\ref{FisherInfoVar}). To that end, we begin with the $Q$ representation~\cite{Hillery1997} for the initial squeezed displaced states $ \prod_{k=0}^{d} |\beta_k;\xi_k\rangle, $ which reads,
\begin{align}
\nonumber Q_0(\bm{r})&=\frac{1}{\pi^{d+1}} \prod_{k=0}^{d} \big| \langle \alpha_k | \beta_k;\xi_k \rangle \big|^2\\
 & =  F(\bm{\beta},\bm{\beta^*})\frac{\exp \left( -\bm{r}^\dg \mathbf{M^\prime} \bm{r} + \bm{r}_b^{\prime\dagger} \bm{r} + \bm{r}^\dagger \bm{r}^\prime_b \right)}{(2\pi)^{d+1} \prod_{k=0}^{d} \cosh|\xi_k|},
\label{Qoriginal}
\end{align}
where $|\alpha_k \rangle$ is a coherent state, 
\begin{align}
\bm{r}=(\bm{\alpha},\bm{\alpha}^*)^T \equiv \left(\alpha_0,\ldots,\alpha_d,\alpha_0^*,\ldots,\alpha_d^* \right)^T,\\
\bm{r}_b^\prime=(\bm{b}^\prime,\bm{b}^{\prime *})^T \equiv \left(b_0^\prime,\ldots,b_d^\prime,b_0^{\prime *},\ldots,b_d^{\prime *} \right)^T,
\end{align}
where $b_j^\prime = \sum_{j=0}^d (\beta_k+\beta_k^* \tanh |\xi_k|)$,
\begin{align}\nonumber
F(\bm{\beta},\bm{\beta^*})=\prod_{k=0}^{d}\exp \left[- \left( |\beta_k|^2+\frac{\tanh|\xi_k|}{2}  \left(\beta_k^{*2}  +  \beta_k^{2}  \right) \right) \right],
\end{align}
\begin{eqnarray}
\nonumber \mathbf{M}^{\prime} = \frac{1}{2}\begin{pmatrix} \mathbf{I} & \mathbf{D} \\ \mathbf{D} & \mathbf{I} \end{pmatrix},
\end{eqnarray}
and 
$\mathbf{D}$ is a diagonal matrix with $D_{j,j}=\tanh|\xi_j|$. The $Q$ representation of the final probe state $\ket{\Psi}$ is then given by $Q(\bm{\alpha}')=|\langle \bm{\alpha}'  \ket{\Psi}|^2,$ where $\bm{\alpha}' = \bm{\mathrm{A}}\bm{\alpha}.$ Our calculation thus exploits the simplicity of applying $\hat{\mathcal{A}}$ on the coherent state basis rather than its conjugate on the squeezed displaced states.
Further simplification is enabled by the passive nature of the transformation $\hat{\mathcal{A}}$ which implies $ |\bm{\alpha}'|^2 = |\bm{\alpha}|^2$ and the $\bm{\phi}-$independence of the QFIM, which can be seen in Eq.~\eqref{FisherInfoVar}. The $Q$ representation of $\ket{\Psi}$ is thus (see App.~\ref{appQ})
\begin{equation}
{Q}(\bm{r})= F(\bm{\beta},\bm{\beta^*})\frac{\exp \left( -\bm{r}^\dg \mathbf{M} \bm{r} + \bm{r}_b^\dagger \bm{r} + \bm{r}^\dagger \bm{r}_b \right)}{(2\pi)^{d+1} \prod_{k=0}^{d} \cosh|\xi_k|},
\label{Qbar}
\end{equation}
with $\bm{r}_b= \left( \bm{b} , \bm{b}^* \right)^T \equiv \left(b_0,\ldots, b_d, b_0^*, \ldots ,b_d^*\right)^T $, and $b_j = \sum_{k=0}^d A_{kj}^* \left( \beta_k +\beta_k^*\tanh|\xi_k| \right).$
The $2(d+1)\times2(d+1)$ matrix $\mathbf{M}$ reads
\begin{eqnarray}
\nonumber \mathbf{M}=\frac{1}{2}\begin{pmatrix} \mathbf{I} & \mathbf{N} \\ \mathbf{N}^\dagger & \mathbf{I} \end{pmatrix},
\end{eqnarray}
with $ \mathbf{N}=\mathbf{A}^\dagger \mathbf{D} \mathbf{A}^*.$
Note that matrix $\mathbf{N}$ is symmetric, i.e., $\mathbf{N}=\mathbf{N}^T$, a fact to be exploited later.

To calculate the QFIM in Eq. (\ref{FisherInfoVar}) using the $Q$ representation of the probe state $\ket{\Psi}$ at hand, we need to recast the expectation values in terms of antinormally ordered operators. These are
$\langle \hat{n}_i \rangle = \langle \hat{a}_i \hat{a}_i^\dagger \rangle - 1$ and $\langle \hat{n}_i \hat{n}_j \rangle = 1+\langle \hat{a}_i \hat{a}_j \hat{a}_i^\dagger \hat{a}_j^\dagger \rangle - \langle \hat{a}_j \hat{a}_j^\dagger \rangle - (1+ \delta_{ij})\langle \hat{a}_i \hat{a}_i^\dagger \rangle$
and can be obtained via a generating function $G(\bm{\mu})$ (see App.~\ref{appGenF})
\begin{equation}
\langle \hat{n}_i \rangle = -1+\left( \frac{\partial}{\partial \lambda_i} \frac{\partial}{\partial \lambda_i^*} \right) G(\bm{\mu}) \Big|_{\bm{\mu}=\bm{0}}
\label{deriv1}
\end{equation}
\begin{eqnarray}
\nonumber \langle \hat{n}_i \hat{n}_j \rangle &=& 1+\Bigg[ \frac{\partial}{\partial \lambda_i} \frac{\partial}{\partial \lambda_i^*} \frac{\partial}{\partial \lambda_j} \frac{\partial}{\partial \lambda_j^*} - \frac{\partial}{\partial \lambda_j} \frac{\partial}{\partial \lambda_j^*} \\
&& - (1+\delta_{i,j})\frac{\partial}{\partial \lambda_i} \frac{\partial}{\partial \lambda_i^*} \Bigg] G(\bm{\mu}) \Big|_{\bm{\mu}=\bm{0}}.
\label{deriv2}
\end{eqnarray}
The generating function, which is based on the $Q$ representation in Eq. (\ref{Qbar}) is given by (see App.~\ref{appGenF}),
\begin{equation}
\label{G2}
G(\bm{\mu})= \exp\left[\frac{ \bm{r}_b^\dagger\mathbf{M}^{-1}\bm{\mu}+\bm{\mu}^\dagger\mathbf{M}^{-1}\bm{r}_b+\bm{\mu}^\dagger\mathbf{M}^{-1}\bm{\mu}
}{4}\right], 
\end{equation}
where $ \bm{\mu}=\left(\lambda_0,\ldots,\lambda_d,\lambda_0^*,\ldots,\lambda_d^* \right)^T$. Note that the derivatives required to calculate the QFIM render the relevant integrals non-Gaussian. Finally, the inverse of $\mathbf{M}$, obtained using Schur's complement, is
\begin{align}
\mathbf{M}^{-1}
=2
\begin{pmatrix}
\mathbf{E} & -\mathbf{N} \mathbf{E}^T\\
-\mathbf{N}^\dagger \mathbf{E} & \mathbf{E}^T
\end{pmatrix},
\label{MmatrixInv}
\end{align}
where $\mathbf{E}=\mathbf{A}^\dagger \mathbf{C} \mathbf{A}$ with $\mathbf{C}$ a diagonal matrix whose non-zero elements read $C_{j,j}= \cosh^2 |\xi_j|$. Note that $\mathbf{E}^{\dagger} = \mathbf{E}$.

By virtue of Eqs. (\ref{deriv1}), (\ref{deriv2}), and (\ref{G2}), the elements $h_{i,j}$ are
\begin{align}
\nonumber h_{i,j} =& 4 \big( (\mathbf{E} \mathbf{N}-\bm{\gamma}\bm{\gamma}^T)\circ(\mathbf{E} \mathbf{N}-\bm{\gamma}\bm{\gamma}^T)^* 
 - (\bm{\gamma}\bm{\gamma}^T)\circ(\bm{\gamma}\bm{\gamma}^T)^*\\\nonumber
& +\frac{1}{4} (\mathbf{E}+\mathbf{E}^*)\circ(\mathbf{E}+\mathbf{E}^*+2\bm{\gamma}\bm{\gamma}^{\dagger}+2\bm{\gamma}^*\bm{\gamma}^T)\\
& -(\mathbf{E}+\bm{\gamma}\bm{\gamma}^{\dagger})\circ\mathbf{I}\big)_{i,j},
\end{align}
where $\bm{\gamma} =  (2\mathbf{E}^* - \mathbf{E}^* \mathbf{N}^{*} -  \mathbf{N}^{*} \mathbf{E}) \bm{b}/2$, $\delta_{i,j}$ is the Kronecker $\delta$ and $\circ$ denotes the Hadamard (entrywise) product. 

Having obtained the general formula for the QFIM, we make two simplifying assumptions to obtain tractable analytical expressions, namely equally squeezed inputs in all the modes ($|\xi_i| = |\xi|, \ \forall i\in\{0,\ldots,d\}$) and an orthogonal interferometer ($\mathcal{A}^\dagger = \mathcal{O}^T \in \mathrm{SO}(d+1)$). What follows in this work relies on these assumptions. For general displacements $\beta_k = x_k + i y_k,$ a straightforward computation leads to a diagonal plus a rank-$1$ matrix,
\begin{equation}
H_{i,j} = \delta_{i,j} h_{i,i} + h_{0,0},
\end{equation}
where $h_{i,i} = 2 \sinh^2 2|\xi| +4 e^{-2 |\xi|} x_i'^2 + 4 e^{2 |\xi|} y_i'^2$ and $h_{0,0} = 2 \sinh^2 2|\xi| +4 e^{-2 |\xi|} x_0'^2 + 4 e^{2 |\xi|} y_0'^2$ 
with $x_i' = \sum_{k=0}^d O_{i,k}^T x_k$ and $y_i' = \sum_{k=0}^d O_{i,k}^T x_k$.
In matrix notation the QFIM reads,
\begin{align}
\mathbf{H} = \mathbf{H}'+h_{0,0} \bm{u}\bm{u}^T,
\end{align}
where $H'_{i,j}=\delta_{i,j} h_{i,i}$ and $u=(1,\ldots,1)^T$. 

We can now bound the total variance of all the parameters, given by $\textrm{Tr}(\mathbf{V}(\bm{\phi})).$ This requires the inverse of the QFIM which, obtained by the Sherman-Morrison formula, is
\begin{equation}
\mathbf{H}^{-1} = \mathbf{H}'^{-1} - \frac{h_{0,0}}{1+h_{0,0} \bm{u}^T \mathbf{H}'^{-1} \bm{u} } \mathbf{H}'^{-1} \bm{u}\bm{u}^T \mathbf{H}'^{-1},
\label{QFIMinv} 
\end{equation}
leading to
\begin{equation}
\label{trace1}
\textrm{Tr}\left(\mathbf{H}^{-1} \right) = \sum_{i=1}^d \frac{1}{h_{i,i}} - \left( \sum_{i=0}^d \frac{1}{h_{i,i}}\right)^{-1} \sum_{i=1}^d \frac{1}{h_{i,i}^2}.
\end{equation}

\section{Simultaneous \textit{\lowercase{vs}} individual phase estimation}\label{sec:simVSind} 

\begin{figure}
    \includegraphics[scale=0.7]{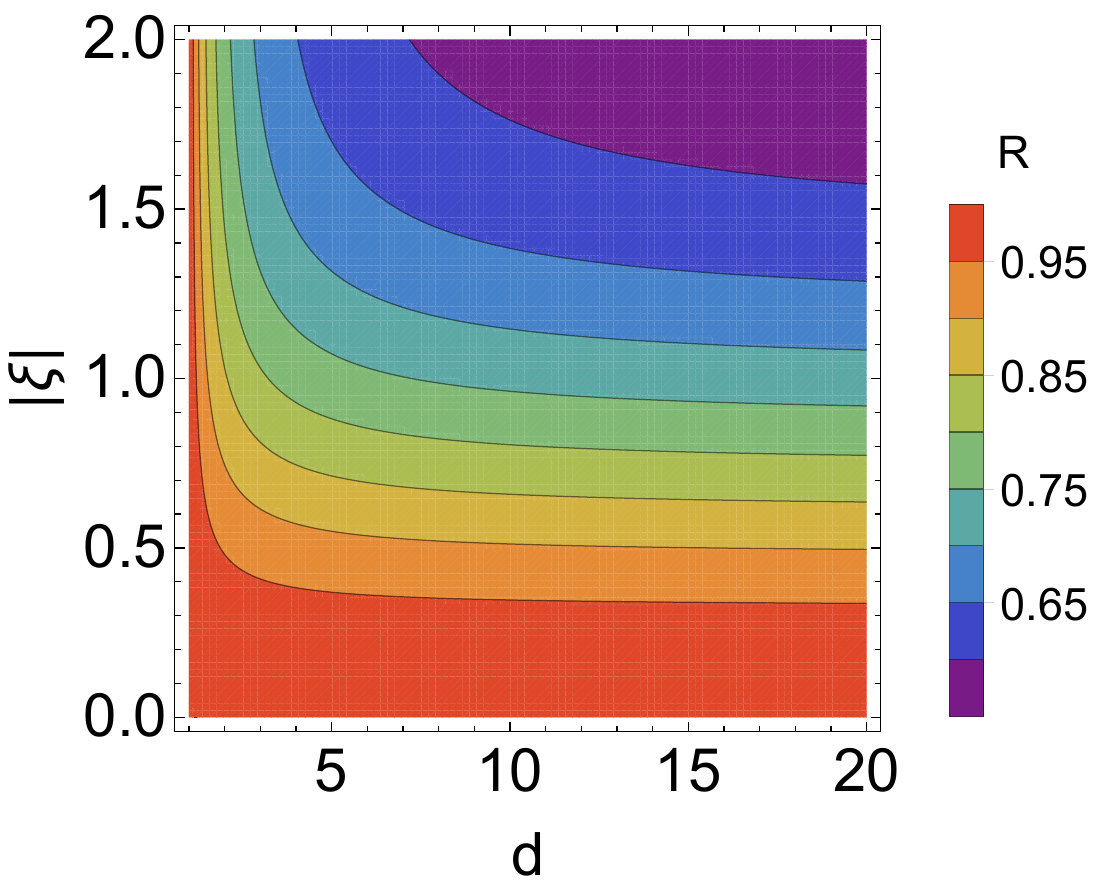}
\caption{$R$ as a function of number of phases and squeezing.}
\label{Rxid}
\end{figure}

The optimal input for estimating the relative phase in a balanced two-mode interferometer is a squeezed state~\cite{Lang2014}. We extend this result within the aforementioned assumptions and prove that for any $d,$ all the energy should go to squeezing for maximal precision in estimation. We do this by first showing that minimising $\textrm{Tr}(\mathbf{H}^{-1})$ is akin to maximising each $h_{i,i}$ independently. Note that $h_{i,i}$ is actually a monotonic function of the fraction of the total energy in displacements, and we show that this quantity is maximum when all the energy is used in squeezing(see App.~\ref{appOpt}). This leads to an optimal QFIM for simultaneous estimation of
\begin{align}
\mathbf{H}_{\textrm{sim}} = 2 (\mathbf{I} + \mathbf{u}\mathbf{u}^T) \sinh^2 2|\xi|
\label{Hsim}
\end{align}
The QFIM $\mathbf{H}_{\textrm{ind}}$ for individual phase estimation comes from the above equation with $d=1.$ We can now compare the quantum limits for the simultaneous estimation of the $d$ phases with their individual estimation for the same expense of energy. The total energy is
$E = \sum_{i=0}^d (x_i^2 + y_i^2) + (d+1) \sinh^2 |\xi| = 2 d \sinh^2|\xi'|,$
where $\xi'$ is the squeezing used for individual estimation. The ratio of the performance of the two estimation strategies is given by
\begin{equation}
R=\frac{\textrm{Tr} \left( \mathbf{H}_{\textrm{sim}}^{-1} \right)} {\textrm{Tr} \left( \mathbf{H}_{\textrm{ind}}^{-1} \right)} = 1-\frac{d-1}{2d}\tanh^2 |\xi|.
\label{ratio}
\end{equation}
In Fig.~\ref{Rxid} the behaviour of $R$ as a function of $|\xi|$ and $d$ is shown.
Since $R \leq 1,$ the simultaneous estimation strategy is superior to the individual estimation strategy. It is also easy to see that $R \geq (1+1/d)/2.$ 
That the ratio $R$ saturates to $1/2$ is unlike the fixed photon number scenario \cite{Humphreys:2013aa} where the limit goes to $0$, although in both cases they fall linearly with $d$.
Possible causes for this are the restriction to Gaussian systems and our assumptions of equal squeezing and orthogonal transformations.
 
\begin{figure}
    \includegraphics[scale=0.6]{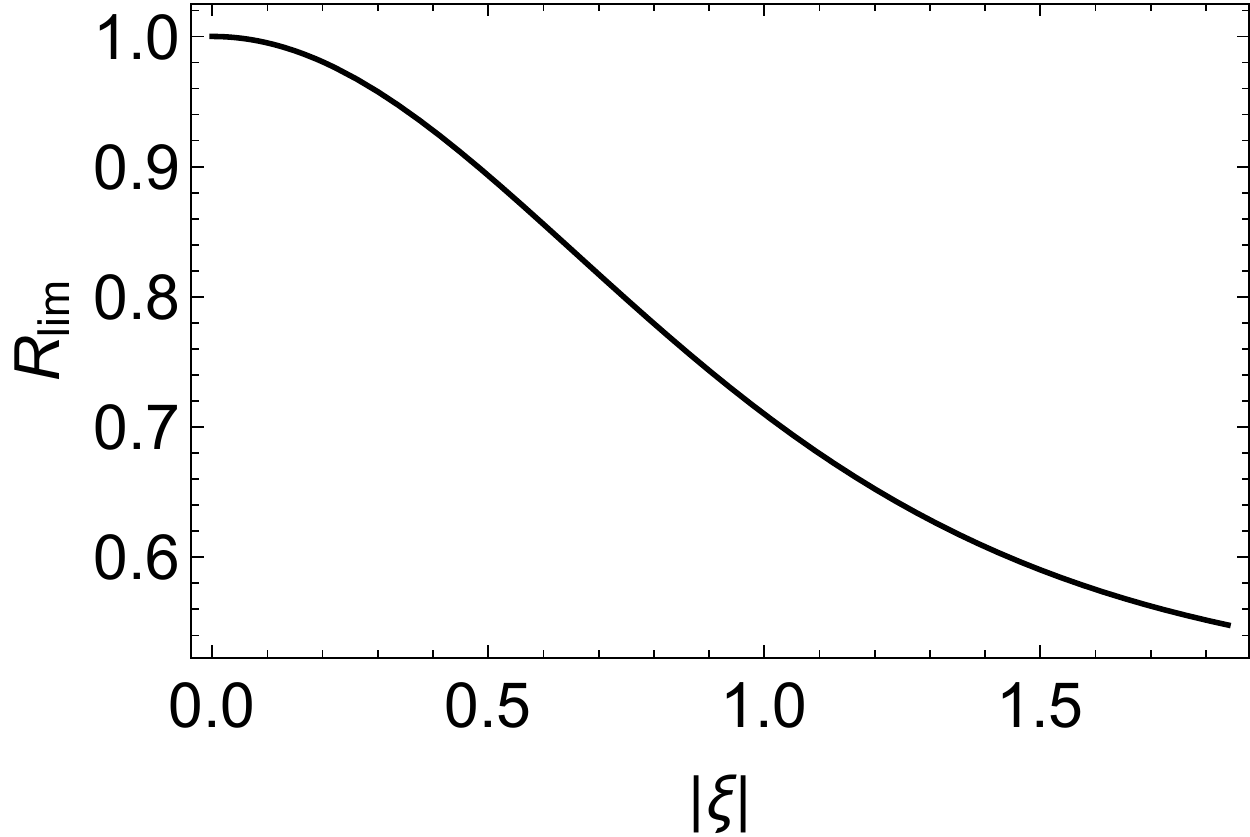}
\caption{$R$ for large number of phases as a function of realistic squeezing. The lower limit is approached quickly for experimentally feasible squeezing.}
\label{plotRlim}
\end{figure}

In the limit of a large number of phase parameters,
\begin{equation}
R_{\textrm{lim}} = \lim_{\ d \to \infty}R = 1-\frac{1}{2}\tanh^2 |\xi|.
\label{Rlim}
\end{equation}
Increasing squeezing is a matter of continuous improvement with state-of-the-art experimental setups, and in Fig.~\ref{plotRlim} we plot $R_{\textrm{lim}}$  for up to $16\ \mathrm{dB}$ squeezing \cite{Schnabel2013}, i.e $|\xi| \approx 1.84$ ($\textrm{dB} = 10 \log_{10} e^{2|\xi|}$). Experimentally, squeezings of $12.7\ \mathrm{dB}$ have been achieved~\cite{eberle2010} along with multimode squeezings of $3.5\ \mathrm{dB}$~\cite{Embrey2015}.

\section{Conclusions}\label{sec:conclusions} We have considered the problem of multiple phase estimation with Gaussian states and have shown that, under some assumptions, the simultaneous estimation of $d$ phases is always superior to the optimum individual estimation strategy. A tentative cause for this improvement is that the simultaneous strategy utilises fewer reference modes, allowing more energy per mode. Our analyses have shown that the larger the variance within a mode the better the estimation. The optimal input states for individual and simultaneous strategies are product squeezed vacuum states and so the distinction boils down to the number of modes; as the simultaneous strategy uses fewer reference modes it allows a larger variance per mode and thus an improved precision. It may be for related reasons that the high-energy limit of the performance ratio of the two strategies coincides with the ratio of the number of modes, $(d+1)/2d$.

It can be noted that these quantum enhancements are obtained from simultaneous estimation without the presence of any quantum entanglement across the modes in the system.
The latter is a consequence of the two assumptions, equal magnitude squeezings and an orthogonal transformation, which we made to obtain analytically tractable expressions.
Nevertheless, this provides---as also claimed in Ref.~\cite{Knott2016}---a possible generalisation of what was known for single phase estimation~\cite{Lang2014,Sahota2015,Friis2015} to multimode interferometry, that modal entanglement is not a crucial resource for quantum-enhanced interferometry.

Our analysis has shown that simultaneous multiple phase estimation is only a factor-of-$2$ better than individual phase estimation using pure Gaussian states. 
This is true for any number $d$ of phases, while with non-Gaussian states the same scenario offers a factor-of-$d$ improvement \cite{Humphreys:2013aa}. 
The limit of a factor-of-$2$ improvement in multimode Gaussian systems as opposed to a factor-of-$d$ with non-Gaussian states seems unique to the multiparameter aspect of the problem.

\begin{acknowledgments} We thank T. Baumgratz and G. Knee for useful discussions. This work was supported by the UK EPSRC (EP/K04057X/2) and the National Quantum Technologies Programme (EP/M01326X/1, EP/M013243/1).
\end{acknowledgments}


\newpage

\onecolumngrid
\appendix
\section*{Appendices}
\renewcommand{\thesubsection}{\arabic{subsection}}

\subsection{Computation of the $Q$ representations}
\label{appQ}
\def\theequation{A\arabic{equation}}
\setcounter{equation}{0}
Initially we consider $d+1$ squeezed displaced states, i.e., $\prod_{k=0}^d | \beta_k; \xi_k \rangle$, where $\xi_k = |\xi_k| e^{i \theta_k}$. From the definition of the $Q$ representation $Q(\bm{r}) = 1/\pi^{d+1} \langle \bm{\alpha} | \hat{\rho} | \bm{\alpha} \rangle$, with $\bm{r}=(\bm{\alpha},\bm{\alpha}^*)^T \equiv \left(\alpha_0,\ldots,\alpha_d,\alpha_0^*,\ldots,\alpha_d^* \right)^T$, one can immediately write,
\begin{eqnarray}
Q_0(\bm{r})&=&\frac{1}{\pi^{d+1}} \prod_{k=0}^{d} \big| \langle \alpha_k | \beta_k;\xi_k \rangle \big|^2.
\label{smQ0}
\end{eqnarray}
The amplitude $\langle \alpha | \beta; \xi \rangle$ can be found as follows,
\begin{eqnarray}
\nonumber \langle \alpha | \beta; \xi \rangle &=& \frac{1}{\sqrt{\cosh|\xi|}} \exp \left( -\frac{|\beta|^2}{2}  -\frac{\beta^*}{2} e^{i \theta} \tanh|\xi| \right) \sum_{n=0}^{\infty} \frac{1}{\sqrt{n!}} H_{n}(\tau) \left( \frac{e^{i \theta} \tanh|\xi|}{2} \right)^{n/2} \langle \alpha | n \rangle \\
\nonumber & = &\frac{1}{\sqrt{\cosh|\xi|}} \exp \left( -\frac{|\beta|^2}{2}  -\frac{\beta^*}{2} e^{i \theta} \tanh|\xi| -\frac{|\alpha|^2}{2}\right) \sum_{n=0}^{\infty} \frac{1}{n!} H_{n}(\tau) \left( \frac{a^{*2}e^{i \theta} \tanh|\xi|}{2} \right)^{n/2}\\
&=&   \frac{1}{\sqrt{\cosh|\xi|}} \exp \left( -\frac{|\alpha|^2}{2} -\frac{|\beta|^2}{2} +\beta \alpha^*  - \frac{1}{2} e^{i \theta} \tanh|\xi|  (\alpha^* - \beta^*)^2  \right),
\label{smAmpl}
\end{eqnarray}
where $H_n (\tau)$ is the Hermite polynomial of the $n$-th order with $\tau = (\beta+\beta^* e^{i \theta} \tanh|\xi|)/ (2 e^{i \theta } \tanh|\xi| )^{1/2}$. We have also used the expansion of a squeezed state in Fock basis \cite{Gong1990} and the Hermite polynomials generating function \cite{Abramowitz1964},
\begin{eqnarray}
\sum_{n=0}^{\infty} \frac{1}{n!} H_n (\tau) \left( \frac{u}{2} \right)^n = \exp \left( 2 \tau u - u^2 \right).
\label{smHermiteGen}
\end{eqnarray}
From Eqs. (\ref{smQ0}) and (\ref{smAmpl}) we write
\begin{eqnarray}
\nonumber Q_0(\bm{r}) &=& \frac{1}{\pi^{d+1} \prod_{k=0}^{d} \cosh|\xi_k|} \prod_{k=0}^{d} \Bigg| \exp\Bigg[ -\frac{|\alpha_k|^2}{2}-\frac{|\beta_k|^2}{2} + \beta_k \alpha_k^*
 -\frac{1}{2} e^{i \theta_k} \tanh|\xi_k|(\alpha_k^*-\beta_k^*)^2 \Bigg] \Bigg|^2\\
 \nonumber &=& \frac{1}{\pi^{d+1} \prod_{k=0}^{d} \cosh|\xi_k|} \\
 && \times \exp\Bigg[ \sum_{k=0}^{d} \left( -|\alpha_k|^2-|\beta_k|^2 + \beta_k \alpha_k^*+ \beta_k^* \alpha_k
 -\frac{1}{2} \tanh|\xi_k| \left( e^{i \theta_k} (\alpha_k^*-\beta_k^*)^2  + e^{-i \theta_k} (\alpha_k-\beta_k)^2 \right) \right) \Bigg].
\label{smQoriginal}
\end{eqnarray}

The state $\prod_{k=0}^d | \beta_k; \xi_k \rangle$ goes through the interferometer denoted as $\hat{\mathcal{A}}^\dagger$ and we take the state $|\Psi \rangle = \hat{\mathcal{A}}^\dagger \prod_{k=0}^d | \beta_k; \xi_k \rangle $. The $Q$ representation of the $|\Psi \rangle $ state is
\begin{eqnarray}
Q(\bm{r}) =  \frac{1}{\pi^{d+1}} \Big| \langle \bm{\alpha} | \hat{\mathcal{A}}^\dagger  \prod_{k=0}^d | \beta_k; \xi_k \rangle \Big|^2.
\label{smQ1}
\end{eqnarray}
It is apparent that it is a lot easier if we act with $\hat{\mathcal{A}}^\dagger $ on the left, i.e., on $\langle \bm{a} |$, that is we consider the transformation $\bm{\alpha'} = \mathbf{A} \bm{\alpha}$ or $\alpha'_k = \sum_{j=0}^d A_{k,j} \alpha_j$. Note that since we consider passive transformations the total energy before and after the interferometer is conserved, i.e., $\sum_{k=0}^d |\alpha'_k|^2 = \sum_{k=0}^d |\alpha_k|^2$.
Applying the transformation and working out Eq. (\ref{smQ1}) a bit we get,
\begin{eqnarray}
\nonumber Q(\bm{r}) &=& \frac{1}{\pi^{d+1} \prod_{k=0}^{d} \cosh|\xi_k|} \exp\Bigg[ - \sum_{k=0}^{d} \left( |\beta_k|^2 + \frac{1}{2} \tanh|\xi_k| \left( e^{i \theta_k} \beta_k^{*2}  + e^{-i \theta_k} \beta_k^2 \right) \right) \Bigg]  \\
\nonumber && \times \exp \Bigg[ -\sum_{k=0}^d \left( |\alpha_k|^2  + \frac{1}{2} \tanh|\xi_k| \left( e^{i \theta_k} \left( \sum_{j=0}^d A^*_{k,j} \alpha_j^* \right)^2  + e^{-i \theta_k} \left( \sum_{j=0}^d A_{k,j} \alpha_j \right)^2 \right) \right) \Bigg]  \\
\nonumber && \times \exp \Bigg[ \sum_{k=0}^d  \beta_k \left(  \sum_{j=0}^d A^*_{k,j} \alpha_j^* + e^{-i \theta_k} \tanh |\xi_k| \sum_{j=0}^d A_{k,j} \alpha_j \right)  \Bigg]  \\
&& \times \exp \Bigg[ \sum_{k=0}^d  \beta_k^* \left(  \sum_{j=0}^d A_{k,j} \alpha_j + e^{i \theta_k} \tanh |\xi_k| \sum_{j=0}^d A^*_{k,j} \alpha_j^* \right)  \Bigg].
\label{smQ2}
\end{eqnarray}
By observing Eq. (\ref{smQ2}) we can write it in a compact form,
\begin{eqnarray}
\nonumber {Q}(\bm{r})&=& F(\bm{\beta},\bm{\beta}^*)\frac{\exp \left( -\bm{r}^\dg \mathbf{M} \bm{r} + \bm{r}_b^\dagger \bm{r} + \bm{r}^\dagger \bm{r}_b \right)}{(2\pi)^{d+1} \prod_{k=0}^{d} \cosh|\xi_k|},
\label{smQbar}
\end{eqnarray}
where
\begin{eqnarray}
F(\bm{\beta},\bm{\beta}^*)=e^{- \sum_{k=0}^{d} \left( |\beta_k|^2+(\frac{1}{2}) \tanh|\xi_k| \left( e^{i \theta_k} \beta_k^{*2}  +  e^{-i \theta_k} \beta_k^{2}  \right) \right)},
\label{smF}
\end{eqnarray}
\begin{eqnarray}
\bm{\beta} &=& \left( \beta_0, \ldots, \beta_d \right),\\
\bm{\beta}^* &=& \left( \beta_0^*, \ldots, \beta_d^* \right),
\end{eqnarray}
\begin{eqnarray}
\bm{r}_b=\left(b_0,\ldots, b_d, b_0^*, \ldots ,b_d^*\right)^T = \left( \bm{b} , \bm{b}^* \right)^T
\end{eqnarray}
with
\begin{eqnarray}
b_j = \sum_{k=0}^d A_{kj}^* \left( \beta_k +\beta_k^* e^{i \theta_k} \tanh|\xi_k| \right).
\label{smb}
\end{eqnarray}
The $2(d+1)\times2(d+1)$ matrix $\mathbf{M}$ reads
\begin{eqnarray}
\mathbf{M}=\frac{1}{2}\begin{pmatrix} \mathbf{I} & \mathbf{N} \\ \mathbf{N}^\dagger & \mathbf{I} \end{pmatrix}
\label{smMmatrix}
\end{eqnarray}
with $ \mathbf{N}=\mathbf{A}^\dagger \mathbf{D} \mathbf{A}^*,$ where $\mathbf{D}$ is a diagonal matrix with $D_{j,j}= e^{i \theta_{j}} \tanh|\xi_j|.$ Note that matrix $\mathbf{N}$ is symmetric, i.e $\mathbf{N}=\mathbf{N}^T$.
Also the matrix $\mathbf{M}$ is Hermitian. In what follows we will need the matrix $\mathbf{M}^{-1}$; to this end we will use Schur's complement \cite{zhang2006}.
We write
\begin{eqnarray}
\mathbf{M}^{-1} = 2 
\begin{pmatrix}
(\mathbf{I} - \mathbf{N} \mathbf{N}^\dagger )^{-1} & - \mathbf{N} (\mathbf{I} - \mathbf{N}^\dagger \mathbf{N}  )^{-1} \\
-\mathbf{N}^\dagger (\mathbf{I} - \mathbf{N} \mathbf{N}^\dagger )^{-1} & (\mathbf{I} - \mathbf{N}^\dagger \mathbf{N}  )^{-1}.
\end{pmatrix}
\label{smMinv1}
\end{eqnarray}
From Eqs. (\ref{smMmatrix}) and (\ref{smMinv1}) it is easy to see that $\mathbf{M}^{-1} \mathbf{M} = \mathbf{I}$. For the Hermitian matrices $\mathbf{N} \mathbf{N}^\dagger$ and $\mathbf{N}^\dagger \mathbf{N}$ we can readily write their diagonalisation (remember that $\mathbf{A}$ is unitary, therefore they diagonalise Hermitian matrices),
\begin{eqnarray}
\label{nnd}\mathbf{N} \mathbf{N}^\dagger &=& \mathbf{A}^\dagger \mathbf{D} \mathbf{D}^\dagger \mathbf{A} \\
\label{ndn}\mathbf{N}^\dagger \mathbf{N} &=& \mathbf{A}^T \mathbf{D} \mathbf{D}^\dagger (\mathbf{A}^T)^\dagger.
\end{eqnarray}
Since $\mathbf{A}^\dagger \mathbf{A} = \mathbf{I}$ and $(\mathbf{A}^T)^\dagger \mathbf{A}^T = \mathbf{I}$ we have

\begin{eqnarray}
\label{sminnd}(\mathbf{I} - \mathbf{N} \mathbf{N}^\dagger )^{-1} &=& \mathbf{A}^\dagger ( \mathbf{I}-\mathbf{D} \mathbf{D}^\dagger)^{-1} \mathbf{A} \\
\label{smindn}(\mathbf{I} - \mathbf{N}^\dagger \mathbf{N} )^{-1} &=& \mathbf{A}^T ( \mathbf{I}-\mathbf{D} \mathbf{D}^\dagger)^{-1} (\mathbf{A}^T)^\dagger.
\end{eqnarray}
The matrix $( \mathbf{I}-\mathbf{D} \mathbf{D}^\dagger)^{-1} \equiv \mathbf{C}$. Since $( \mathbf{I}-\mathbf{D} \mathbf{D}^\dagger)^{-1}$ is a diagonal matrix, the matrix $\mathbf{C}$ is easily found to be the diagonal matrix whose non-zero elements read $C_{j,j}= \cosh^2 |\xi_j|$. Therefore from Eqs. (\ref{smMinv1}), (\ref{sminnd}) and (\ref{smindn}) we write
\begin{eqnarray}
\mathbf{M}^{-1}
=2
\begin{pmatrix}
\mathbf{E} & -\mathbf{N} \mathbf{E}^T\\
-\mathbf{N}^\dagger \mathbf{E} & \mathbf{E}^T
\end{pmatrix},
\label{smMmatrixInv}
\end{eqnarray}
where $\mathbf{E}=\mathbf{A}^\dagger \mathbf{C} \mathbf{A}$. 

Let us now prove that the matrix $\mathbf{M}$ is not only Hermitian but also positive semidefinite and therefore can be used in the next section as a complex covariance matrix. We will calculate the (real) eigenvalues $\sigma$ of the (Hermitian) matrix $\mathbf{M}$. The characteristic polynomial reads
\begin{eqnarray}
\det
(\mathbf{M} - \sigma\mathbf{I} )= \det  \begin{pmatrix} \left(\frac{1}{2} - \sigma \right)\mathbf{I} & \frac{1}{2} \mathbf{N}\\
 \frac{1}{2} \mathbf{N}^\dagger & \left(\frac{1}{2} -  \sigma\right) \mathbf{I} \end{pmatrix} =0
\label{smDet}
\end{eqnarray} 
Since the blocks in Eq. (\ref{smDet}) are square and $\mathbf{N}^\dagger$  commutes with $\left( \frac{1}{2} -\sigma \right)\mathbf{I}$, from \cite{silvester2000} we can write
\begin{eqnarray}
\det
(\mathbf{M} - \sigma \mathbf{I} )= \det \left( \left(\frac{1}{2} -  \sigma\right)^2 \mathbf{I} - \frac{1}{4} \mathbf{N} \mathbf{N}^\dagger  \right) = 0.
\label{smDet2}
\end{eqnarray}
By virtue of Eq. (\ref{nnd}) and the facts that $\mathbf{A}$ is unitary and $\mathbf{M} - \sigma \mathbf{I}$ is Hermitian, and by substituting the elements of the diagonal matrices $\mathbf{D}$ and $\mathbf{D}^\dagger$, we can write
\begin{eqnarray}
 \det(\mathbf{M} - \sigma \mathbf{I} ) = \det \left( \left(\frac{1}{2} -  \sigma\right)^2 \mathbf{I} - \frac{1}{4} \mathbf{D} \mathbf{D}^\dagger  \right)
= \prod_{i=0}^{d} \left[ \left(\frac{1}{2} -  \sigma_i \right)^2-\frac{1}{4} \tanh^2 |\xi_i| \right] = 0.
\label{smDet3}
\end{eqnarray}
From Eq. (\ref{smDet3}) we readily find
\begin{eqnarray}
\sigma_i  = \frac{1}{2} \left( 1 \mp \tanh |\xi_i | \right) \ge 0
\end{eqnarray}

\subsection{Generating function and mean values}
\label{appGenF}
We introduce the generating function $G(\bm{\mu})$,
\begin{eqnarray}
G(\bm{\mu})= \int \mathrm{d}\bm{r}  Q(\bm{r}) \exp\left( \sum_{j=0}^d \lambda_j \alpha^*_j + \sum_{j=0}^d \lambda_j^* \alpha_j \right),
\label{smGenerating1}
\end{eqnarray}
where $ \bm{\mu}=\left(\lambda_0,\ldots,\lambda_d,\lambda_0^*,\ldots,\lambda_d^* \right)^T$. The $\lambda$'s are the so-called sources \cite{zeidler2009}, nothing else than some helping parameters when it comes to calculating somewhat difficult integrals \cite{nahin2014}. The word \emph{sources} comes from the fact that some linear terms are added into the exponential. Sometimes this is referred to as \emph{Feynman's favourite trick}.
It is not difficult to see that the integral in Eq. (\ref{smGenerating1}) is just a Gaussian integral and is therefore easy to be calculated. Also observe that when we hit Eq. (\ref{smGenerating1}) with derivatives with respect to $\lambda$'s at $\bm{\mu}=\bm{0}$, we get expectation values of combinations of $\hat{a},\ \hat{a}^\dagger$, that justifies the name \emph{generating function}. This is exactly what we need in order to calculate the QFIM for pure states. 
Since we use the $Q$ representation formalism we must calculate expectation values in terms of the mean values of antinormally ordered operators, i.e., all creation operators should be on the right,
\begin{eqnarray}
\label{smMean1} \langle \hat{n}_i \rangle &=& \langle \hat{a}_i \hat{a}_i^\dagger \rangle - 1,\\
\langle \hat{n}_i \hat{n}_j \rangle &=& \langle \hat{a}_i \hat{a}_j \hat{a}_i^\dagger \hat{a}_j^\dagger \rangle
- \langle \hat{a}_i \hat{a}_i^\dagger \rangle- \langle \hat{a}_j \hat{a}_j^\dagger \rangle
- \langle \hat{a}_i \hat{a}_i^\dagger \rangle \delta_{ij} + 1,
\label{smMean2}
 \end{eqnarray}
 where we have used $[\hat{a},\hat{a}^\dagger]=1$. From Eqs. (\ref{smGenerating1}), (\ref{smMean1}) and (\ref{smMean2}) it is not difficult to see that,
  \begin{eqnarray}
\langle \hat{n}_i \rangle = \left( \frac{\partial}{\partial \lambda_i} \frac{\partial}{\partial \lambda_i^*} \right) 
G(\bm{\mu}) \Bigg|_{\bm{\mu}=\bm{0}} - 1,
\label{smDeriv1}
\end{eqnarray}

 \begin{eqnarray}
 \langle \hat{n}_i \hat{n}_j \rangle =\Bigg[ \frac{\partial}{\partial \lambda_i} \frac{\partial}{\partial \lambda_i^*} \frac{\partial}{\partial \lambda_j} \frac{\partial}{\partial \lambda_j^*} - 
 (1+\delta_{ij})\frac{\partial}{\partial \lambda_i} \frac{\partial}{\partial \lambda_i^*} - \frac{\partial}{\partial \lambda_j} \frac{\partial}{\partial \lambda_j^*} \Bigg] G(\bm{\mu}) \Bigg|_{\bm{\mu}=\bm{0}} + 1.
\label{smDeriv2}
\end{eqnarray}
So, we have transformed the problem of calculating a non-Gaussian integral (when calculating the mean photon number for example) into one of calculating a Gaussian integral and its derivatives up to fourth order.

In Eq. (\ref{smGenerating1}) by $\mathrm{d}\bm{r}$ we denote integration over all $\Re{\alpha}$ and $\Im{\alpha}$. However, we find it more convenient to calculate the integral over $\alpha$ and $\alpha^*$. To this end we will need the Jacobian for the transformation $(\Re{\alpha},\ \Im{\alpha}) \rightarrow (\alpha, \alpha^*)$, which reads $1/2^{d+1}$.
By doing the Gaussian integral of Eq. (\ref{smGenerating1}) we find the generating function,
\begin{eqnarray}
G(\bm{\mu})=F(\bm{\beta},\bm{\beta}^*)\frac{\exp\left[ (\bm{r}_b^\dagger + \bm{\mu}^\dagger )  \mathbf{M}^{-1}  (\bm{r}_b + \bm{\mu} ) \right]}{2^{d+1} \det{\mathbf{M}} \prod_{k=0}^d \cosh|\xi_k|},
\label{smG1}
\end{eqnarray}
where $F(\bm{\beta},\bm{\beta}^*)$ was defined in Eq. (\ref{smF}).

We can simplify the generating function even more by noting that $G(\bm{\mu}=\bm{0})=1$ since this is simply the integration of the $Q$ representation over all phase space, i.e., this is just the normalization to $1$ of the $Q$ quasiprobability distribution. Therefore we get
\begin{eqnarray}
G(\bm{\mu})= \exp \left[ \frac{1}{4} \left( \bm{r}_b^\dagger\mathbf{M}^{-1}\bm{\mu}+
\bm{\mu}^\dagger\mathbf{M}^{-1}\bm{r}_b+
\bm{\mu}^\dagger\mathbf{M}^{-1}\bm{\mu} \right)
\right].
\label{smG2}
\end{eqnarray}

Now the job is straightforward, easy, and boring; by carefully performing the derivatives of Eqs. (\ref{smDeriv1}) and (\ref{smDeriv2}) one finds the matrix elements $h_{i,j}$ and therefore the QFIM found in the main body of the text.

\subsection{Optimization}
\label{appOpt}
We have given the expression for $\trace{\mathbf{H}^{-1}}$ in terms of the elements $h_{i,i}$ under the assumptions that we have an equal squeezing in each mode and that the unitary transform is an orthogonal transform as
\begin{align}\label{eq:QFIhjj}
\trace{\mathbf{H}^{-1}}= \sum_{i=1}^d \frac{1}{h_{i,i}} - \left( \sum_{i=0}^d \frac{1}{h_{i,i}}\right)^{-1} \sum_{i=1}^d \frac{1}{h_{i,i}^2}.
\end{align}
We can rewrite $h_{i,i} = 2 \sinh^2 2|\xi| +4 e^{-2 |\xi|} x_i'^2 + 4 e^{2 |\xi|} y_i'^2$ in terms of some $E_j=\sinh^2|\xi|+x_j'^2+y_j'^2$, and $E_{\gamma_j}=x_j'^2+y_j'^2$ and $\theta_{\gamma_j}=\cos^{-1}\left(\frac{x_j'}{\sqrt{E_{\gamma_j}}}\right)$.  Under this parameterisation the energy constraint becomes $\sum_{j=0}^d E_j=E_{\mathrm{Tot}}$\footnote{The energy constraint was originally written in terms of the displacements before the passive unitary, however as it is passive $\sum_{i=0}^d x_i^2+y_i^2=\sum_{i=0}^d x_i'^2+y_i'^2$ allows us to say that the total energy due to displacements displacements before the unitary is equal to the total energy of the displacements after the unitary}.  We thus write $h_{j,j}=h_{j,j}(E_j,E_{\gamma_j},\theta_{\gamma_j})$ and can now extremise $\trace{\mathbf{H}^{-1}}$ over $E_{\gamma_j}$ and $\theta_{\gamma_j}$ without needing to construct a Lagrangian problem (as the only constraint on $E_{\gamma_j}$ and $\theta_{\gamma_j}$ is $0\leq E_{\gamma_j}\leq E_j$). We now consider what we need to solve in order to extremise $\trace{\mathbf{H}^{-1}}$ with respect to $m_j=E_{\gamma_j},\theta_{\gamma_j}$ for $j\neq 0$~\footnote{It is clear from Eq.~\eqref{eq:QFIhjj} that $\trace{\mathbf{H}^{-1}}$ is minimised when $h_{0,0}$ is maximised.}
\begin{align}\nonumber
\frac{\partial \trace{\mathbf{H}^{-1}}}{\partial m_j}&=
\frac{\partial h_{j,j}}{\partial m_j}
\left[
-\frac{1}{h_{j,j}^2}
+\frac{2}{h_{j,j}^3}\left(\sum_{i=0}^d\frac{1}{h_{i,i}}\right)^{-1}
-\frac{1}{h_{j,j}^2}\left(\sum_{i=0}^d\frac{1}{h_{i,i}}\right)^{-2}\sum_{i=1}^d\frac{1}{h_{i,i}^2}
\right]\\
&=
-\frac{\partial h_{j,j}}{\partial m_j}\frac{1}{h_{j,j}^2}\left(\sum_{i=0}^d\frac{1}{h_{i,i}}\right)^{-2}
\left[
\left(\sum_{i=0}^d\frac{1}{h_{i,i}}\right)^2
-\frac{2}{h_{j,j}}\sum_{i=0}^d\frac{1}{h_{i,i}}
+\sum_{i=1}^d\frac{1}{h_{i,i}^2}
\right]
\end{align}
We first note that the terms in the square brackets can be rewritten (for $j\neq 0$) as
\begin{align}
\left(\sum_{i=0,i\neq j}^d\frac{1}{h_{i,i}}\right)^2+\frac{2}{h_{j,j}}\sum_{i=0}^d\frac{1}{h_{i,i}}
-\frac{2}{h_{j,j}}\sum_{i=0}^d\frac{1}{h_{i,i}}
+\sum_{i=1,i\neq j}^d\frac{1}{h_{i,i}^2}
\end{align}
The middle two terms cancel and the remaining terms are clearly positive.  Thus we may freely conclude that
\begin{align}
\frac{\partial \trace{\mathbf{H}^{-1}}}{\partial m_j}=-\kappa \frac{\partial h_{j,j}}{\partial m_j},\kappa>0
\end{align}
Namely, to extremise $h_{j,j}$ with respect to $m_j$ is to extremise $\trace{\mathbf{H}^{-1}}$ with respect to $m_j$; furthermore as $\kappa>0$ if a change in $m_j$ increases $h_{j,j}$ then it necessarily decreases $\trace{\mathbf{H}^{-1}}$.  We now therefore turn our attention to the maximisation of $h_{j,j}$ with respect to $E_{\gamma_j}$ and $\theta_{\gamma_j}$,
\begin{align}
h_{j,j}=&
4\left(-E_{\gamma_j}+2E_j(1+E_j-E_{\gamma_j})\right)+8E_{\gamma_j}\sqrt{(E_j-E_{\gamma_j})(1+E_j-E_{\gamma_j})}\cos 2\theta_{\gamma_j}
\end{align}
No further mathematics is required to see that $h_{j,j}$ is maximised with respect to $\theta_{\gamma_j}$ by $\theta_{\gamma_j}=0$, which reduces the problem to
\begin{align}
h_{j,j}=&
4\left(-E_{\gamma_j}+2E_j(1+E_j-E_{\gamma_j})\right)+8E_{\gamma_j}\sqrt{(E_j-E_{\gamma_j})(1+E_j-E_{\gamma_j})},\\\label{eq:hjjderivative}
\frac{\partial h_{j,j}}{\partial E_{\gamma_j}}=&
-4-8E_j+8\sqrt{(E_j-E_{\gamma_j})(1+E_j-E_{\gamma_j})}+4E_{\gamma_j}\frac{-1-2E_j+2E_{\gamma_j}}{\sqrt{(E_j-E_{\gamma_j})(1+E_j-E_{\gamma_j})}}=0
\end{align}
We can then solve Eq.~\eqref{eq:hjjderivative} to find the solutions $\sqrt{(E_j-E_{\gamma_j})(1+E_j-E_{\gamma_j})}=-E_{\gamma_j}$ and $\sqrt{(E_j-E_{\gamma_j})(1+E_j-E_{\gamma_j})}=E_j+E_{\gamma_j}+\half$.  Both of these entail $E_{\gamma_j}$ to lie outside of $0\leq E_{\gamma_j}\leq E_j$ (the former obviously so, the latter solutions requires the similarly unacceptable $E_{\gamma_j}=-\frac{1}{8(1+2E_j)}$).  To this end there are no extrema within the allowed values of $E_{\gamma_j}$ instead $h_{j,j}$ is monotonic within those values.  To this end we consider the extreme cases, $E_{\gamma_j}=0$ and $E_{\gamma_j}=E_j$, which yield respectively $h_{j,j}=8E_j(E_j+1)$ and $h_{j,j}=4E_j$.  $E_{\gamma_j}=0$ could have been expected to yield the superior solution as $E_{\gamma_j}=E_j$ corresponds to the use of a coherent state.
We are now left to optimise $\trace{\mathbf{H}^{-1}}$ over $\{E_j\}$ subject to $\sum_{j=0}^d E_j=E_{\mathrm{Tot}}$, however as $E_{\gamma_j}=\sinh^2 |\xi_j|$ we have previously assumed $|\xi_j|=|\xi|,\forall j\in\{0,\ldots,d\}$ which takes us to $E_j=\frac{E_{\mathrm{Tot}}}{d+1}$; this leads us to the optimal QFIM, 
\begin{align}\nonumber
\mathbf{H}_{\mathrm{sim}}&=2(\mathbf{I}+\mathbf{u}\mathbf{u}^T)\sinh^2 2|\xi|\\
&=8(\mathbf{I}+\mathbf{u}\mathbf{u}^T)\frac{E_{\mathrm{Tot}}(d+1+E_{\mathrm{Tot}})}{(d+1)^2}.
\end{align}

\twocolumngrid
\bibliography{BibliographyMetrology_2,BibliographySM}

\end{document}